\documentclass[prl,twocolumn,superscriptaddress]{revtex4-2}

\usepackage{graphicx}  
\usepackage{amsmath}   
\usepackage{amssymb}   
\usepackage{hyperref}  
\usepackage{natbib}    

\usepackage[normalem]{ulem}
\usepackage{lineno}
\usepackage[usenames,dvipsnames]{xcolor}
\modulolinenumbers[5]

\begin{document}

\title{Segregation in binary mixture with differential contraction among active rings}

\author{Emanuel F. Teixeira}
\email{emanuel.teixeira@ufrgs.br}
\affiliation{Instituto de Física, Universidade Federal do Rio Grande do Sul, CP 15051, CEP 91501-970 Porto Alegre - RS, Brazil}

\author{Carine P. Beatrici}
\email{carine@if.ufrgs.br}
\affiliation{Instituto de Física, Universidade Federal do Rio Grande do Sul, CP 15051, CEP 91501-970 Porto Alegre - RS, Brazil}

\author{Heitor C. M. Fernandes}
\email{heitor.fernandes@ufrgs.br}
\affiliation{Instituto de Física, Universidade Federal do Rio Grande do Sul, CP 15051, CEP 91501-970 Porto Alegre - RS, Brazil}
\author{Leonardo G. Brunnet}
\email{leon@if.ufrgs.br}
\affiliation{Instituto de Física, Universidade Federal do Rio Grande do Sul, CP 15051, CEP 91501-970 Porto Alegre - RS, Brazil}

\date{\today}

\begin{abstract}
Cell cortex contraction is essential for shaping cells, enabling movement, ensuring proper division, maintaining tissue integrity, guiding development, and responding to mechanical signals — all critical for the life and health of multicellular organisms. Differential contractions in cell membranes, particularly when cells of different types interact, play a crucial role in the emergence of segregation. In this study, we introduce a model where rings composed of active particles interact through differential membrane contraction within a specified cutoff distance. We demonstrate that segregation arises solely from differential contraction, with the activity of the rings functioning similarly to an effective temperature.
Additionally, we observed that segregation involves cluster fusion-diffusion process. However, the decay exponent of the segregation parameter we found is close to $\lambda \sim -1/3$, which differs from the $\lambda \sim -1/4$ predicted by previous theoretical approaches and simulations.

\end{abstract}

\maketitle

In cellular systems made up of different species, spontaneous sorting is a common emergent behavior.
During embryonic development, cells undergo differentiation, which leads to the spontaneous segregation of cell types in tissue formation. Based on strong experimental evidence \cite{harrison1910outgrowth,moscona1952cell,trinkaus1955differentiation,weiss55,weiss1960reconstitution} and motivated by physical systems like binary mixtures, where spontaneous separation of two liquids is observed,  Steinberg proposes \cite{steinberg1963reconstruction} that the general mechanism for cell segregation lies in the difference in adhesion between cells of  different  types.
This proposal became known in the literature as the Differential Adhesion Hypothesis (DAH) or Steinberg Hypothesis.

Harris \cite{harris1976} questions the foundation of the DAH hypothesis, highlighting that mere maximization of intercellular adhesion does not necessarily lead to the observed effects of cell sorting. He suggested that within a heterogeneous cell aggregate, variation in surface contraction, driven by the active regulation of the acto-myosin cortex, could be the driving force behind tissue engulfment and cell sorting in vivo.
The surface contraction of a particular cell appearing more pronounced when it comes into contact with a cell of a different histological type. 
Besides that, cells from different tissues exert different surface contraction when they contact the medium. This sorting mechanism was named Differential Surface Contraction Hypothesis (DSCH). 

In line with Harris's ideas, Brodland \cite{brodland2002differential} successfully described segregation using finite element simulations that incorporated adhesion and surface contractions, both contributing to the total interfacial tension, a mechanism known as differential interfacial tension hypothesis (DITH).
In fact, these hypotheses have been tested using various other extended numerical models, including Cellular Potts \cite{graner1992simulation,kafer2007cell,krieg2008tensile,nakajima2011kinetics,canty2017sorting,durand2021large} and Vertex models\cite{barton2017,sussman2018soft,wolff2019adapting,krajnc2020solid}. 
All theses models define an effective interfacial tension between different units but do not explicitly separate adjacent cell membranes or detail the forces involved, such as adhesion and surface contraction. 
However, from an experimental perspective, the role of membrane fluctuations was clearly highlighted by Mombach \cite{mombach95}, and the distinction between adhesion and cortical tension was emphasized in the work of Krieg et al.\cite{krieg2008tensile} and Manning et al. \cite{manning2010}.

In this work, we present a model of active rings that interact through differential membrane contraction when within a specific cutoff distance. By utilizing two interacting membranes, this model provides a more detailed representation of biological processes, offering insights into how different layers interact and differentiating the roles of fluctuations, adhesion, and cortex contraction. To our knowledge, this is the first time Harris's mechanism has been simulated in isolation.

\begin{figure}[h]
\centering
\includegraphics[width=8.6cm]{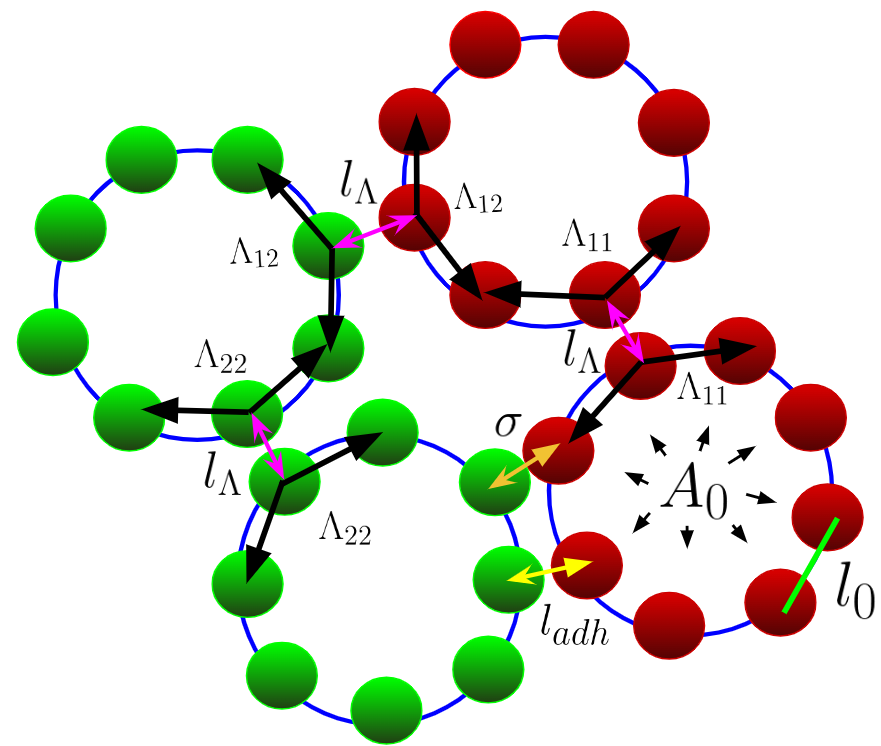}
\caption{
Representation of the active ring model used in this work. Differential contractions occur between active rings when two particles from different rings come within a distance $l_\Lambda$. At this proximity, they experience a force directed along the springs, pushing the particles toward the center of their respective rings. This interaction induces membrane contraction (indicated by black arrows), and the extent of this contraction depends on the type of ring to which the particle belongs.
}
\label{rings_scheme}
\end{figure}

Boromand and collaborators \cite{boromand2018jamming,treado2021bridging} introduced a ring system composed of passive particles connected by springs. 
Building on this model, we incorporated active properties into the particles forming the ring in previous articles \cite{teixeira2021single,ourique2022modelling}. We model a 2D system with $N$ rings, each representing a cell formed by $n$ active particles (see Fig. \ref{rings_scheme}).
The system is set up as a binary mixture of active rings confined in a circular arena with repulsive walls and radius $R_0$. 
The set of coupled overdamped equations governing the dynamics of each particle are: $\mathbf{\dot{r}}_{i,j} = v_{0}\;\mathbf{n}_{j} -\mu\,\mathbf{F}_{i,j}$ and $\mathbf{\dot{n}}_{j} = \sqrt{2\,D_{R}}\; \boldsymbol{\xi}_{j}\times \mathbf{n}_{j}$,
where $\mathbf{r}_{i,j}$ denotes the position of the  $i$-th particle within the $j$-th ring at time $t$, $\mu$ represents its mobility and $v_{0}$ is the magnitude of active velocity with its orientation given by $\mathbf{n}_{j}$. 
The second term, $\mathbf{F}_{i,j} = -\boldsymbol{\nabla}_{i,j} E$ represents the total force acting on the $i$-th particle within the $j$-th ring.
The direction of the active force, described by unit vector $\mathbf{n}_{j}$, experiences angular Gaussian white noise $\boldsymbol{\xi}_{j} = \xi_{j}\hat{e}_{z}$ with correlation $\left \langle \xi_{j}(t_{1})\xi_{k}(t_{2}) \right \rangle = \delta_{jk}\delta (t_{1}-t_{2})$. The noise term $D_{R}$ represents the rotational diffusion coefficient, which defines a characteristic timescale given by $\tau_{R} = 1/D_{R}$.

The energy function has contributions from $(i)$ a perimeter energy (springs connecting ring neighboring particles), $(ii)$ area conservation, $(iii)$ contact-dependent contraction term, $(iv)$ core repulsion among non-neighboring particles from any ring and inter-cellular adhesion, 
\begin{eqnarray}
\label{2}
E &=& \sum_{j=0}^{N} \left\{\frac{\epsilon_{P}}{2}  \sum_{i=0}^{n} \left( \frac{|\vec{l}_{i,j}|}{l_{0}} - 1 \right)^{2} + \frac{\epsilon_{A}}{2}  \left( \frac{A_{j}}{A_{0}} - 1 \right)^{2} \right. \nonumber \\  
&+& \left. \sum_{i=0}^{n} \Lambda_{\alpha \beta} |\vec{l}_{i,j}| \right\} + \frac{\epsilon_{c}}{2} \sum_{r_{ik} \leq  \sigma} \left( \frac{r_{ik}}{\sigma} - 1 \right)^{2} \nonumber \\  
&+& \frac{\epsilon_{adh}}{2} \sum_{\sigma < r_{ik} \leq  l_{adh}} \left( \frac{r_{ik}}{\sigma} - 1 \right)^{2},
\end{eqnarray}
where $\vec{l}_{i,j} = \vec{r}_{i,j} - \vec{r}_{i-1,j}$ is the vector connecting consecutive particles in ring $j$, $\epsilon_{P}$ is the elastic energy of the spring controlling perimeter fluctuations, and $l_{0}$ is the equilibrium distance in the ring. 
The elastic energy related to area control is $\epsilon_{A}$, $A_{j}$ is the inner area of the $j$-th ring (not considering the area of particles), and $A_{0}$ is the equilibrium area. The last two terms in Eq. \ref{2} represent the core repulsion between non-neighboring particles  and the adhesion between particles of different rings, respectively. Parameter $\epsilon_{c}$ denotes the characteristic energy of the core repulsion interaction, while $r_{ik}$ is the distance between particles $i$ and $k$. The equilibrium cut-off distance, $\sigma$, effectively defines the particle diameter. The adhesion energy and interaction distance are characterized by $\epsilon_{adh}$ and $l_{adh}$, respectively.

\begin{figure*}[!ht]
    \centering
 \includegraphics[width=0.5\textwidth]{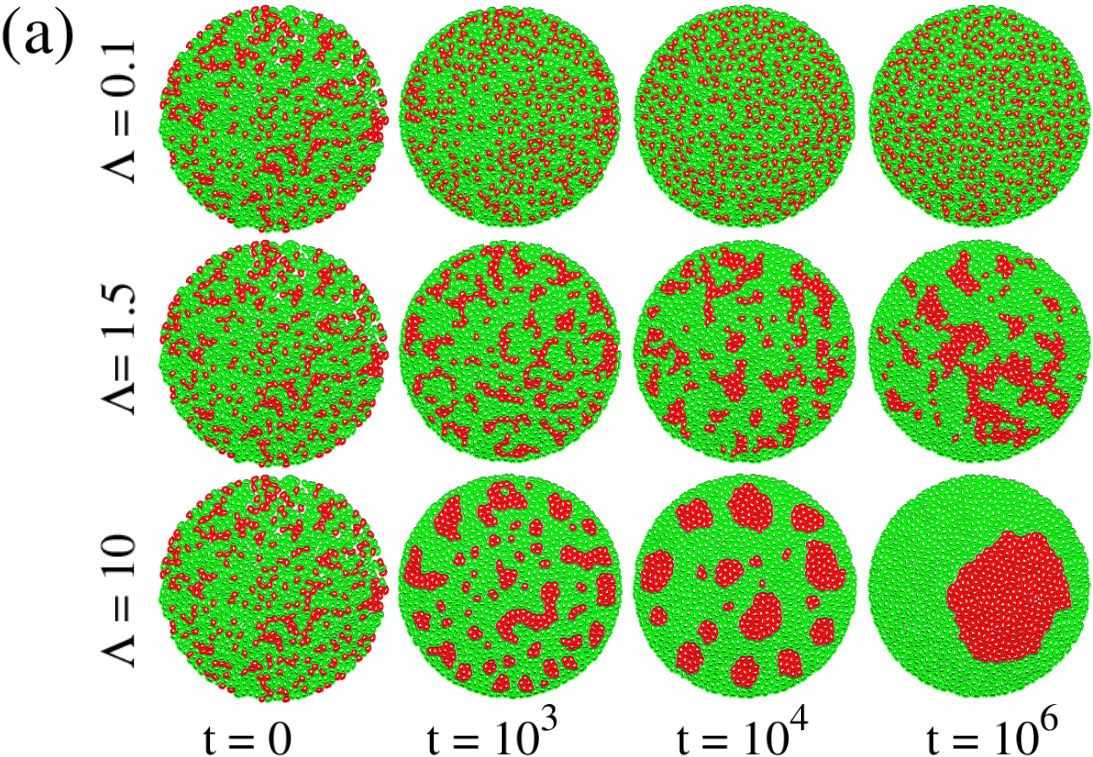}
  \vspace{0.1cm} \hspace{0.5cm} 
   \includegraphics[width=0.35\textwidth]{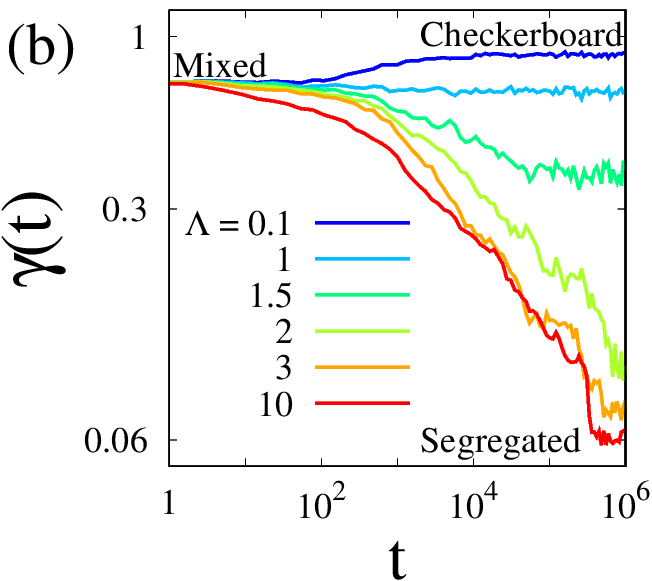}
	\caption{(a) Simulation snapshots of a binary mixture of rings with differential membrane contraction for various values of $\Lambda$. For $\Lambda=10$ (bottom), a stable segregation state emerges, where all clusters eventually merge into a final configuration with a single cluster of the minority type (red rings). For $\Lambda=1.5$ (center), the system begins to segregate, but cluster formation remains unstable, with clusters continuously breaking apart even at long times. For $\Lambda=0.1$ (top), the system transitions from an initially mixed state to a configuration where rings prefer contact with those of a different type, forming a checkerboard-like pattern. 
b) Time series of the segregation parameter for various differential contractions $\Lambda$. The system begins in a mixed configuration with 30\% red rings and 70\% green rings. For $\Lambda<1$, rings energetically prefer contact with a different type, leading to the formation of a checkerboard pattern and an increase in $\gamma$. For $\Lambda>1$, rings settle into a configuration where they are in contact with rings of the same type, resulting in a decrease of the interface and, consequently, in the $\gamma$ value.
 }
 \label{SNAPSHOTS_and_GAMMA_TIME-SERIE}
\end{figure*}

A key distinction between the ring model and other extensive models (Finite Elements \cite{brodland2002differential} and Vertex-Voronoi model  \cite{bi2015density,bi2016motility,barton2017,sussman2018soft,wolff2019adapting,krajnc2020solid}) lies in the rings contact interface, which comprises two contact membranes. 
This bring us to the third term in Eq. \ref{2} which incorporates differential surface contraction. In this term, parameter $\Lambda_{11}$ represents the line tension between particles belonging to type 1 rings (red rings in Fig. \ref{rings_scheme}), $\Lambda_{22}$ represents the line tension between type 2 rings and $\Lambda_{12}=\Lambda_{21}$ represents the line tension between rings of different types. All these tensions acting when they are within a cutoff distance $l_{\Lambda}$.
A central parameter in this work is $\Lambda = \Lambda_{12}/\bar{\Lambda}$, with  $\bar{\Lambda} \equiv (\Lambda_{11}+\Lambda_{22})/2$.
Fig. \ref{rings_scheme} shows a schematic representation of the active ring system, illustrating  the involved tensions. 
 
To handle the interaction with the walls, which can be seen as the medium in our context, we specify that the particles of a ring increase their contraction tension once they reach a distance $\sigma/2$ from the wall. 
This prevents the rings from preferring to accumulate on the wall. 
The contraction between rings of type 1 (red) and the ``medium'' (wall) is defined as $\Lambda_{1M}/\bar{\Lambda} = 12$, while for rings of type 2 (green), we use $\Lambda_{2M}/\bar{\Lambda} = 0$. 
This choice  satisfies the criterion $\Lambda_{12}< \Lambda_{1M}-\Lambda_{2M}$ (valid for all values of $\Lambda_{12}$ in this work) for the engulfment of one type (red) by the other (green) \cite{brodland2002differential,barton2017}.

We measure spatial coordinates and time in units of $l_{0}$ and $\tau_{R}$, respectively. 
As a result, the model equations may be written in terms of the Péclet number~\cite{fily2012athermal,fily2014freezing}, defined as $Pe \equiv \frac{v_0 \tau_{R}}{l_{0}}$ , which controls the level of activity. Details of the numerical integration method and the remaining parameter values are provided in the Supplemental Material.

We begin our simulations with an initial random distribution of rings (see Fig. \ref{SNAPSHOTS_and_GAMMA_TIME-SERIE}a). For $\Lambda > 1$, we observe a spontaneous segregation process among the rings, with larger $\Lambda$ values resulting in a smaller interface between ring types, indicating improved segregation. This is illustrated at the bottom of Fig. \ref{SNAPSHOTS_and_GAMMA_TIME-SERIE}a, where $\Lambda = 10$. For $\Lambda = 1.5$ (middle of Fig. \ref{SNAPSHOTS_and_GAMMA_TIME-SERIE}a), the system still segregates, but a larger asymptotic interface remains. In contrast, for $\Lambda < 1$ (top of Fig. \ref{SNAPSHOTS_and_GAMMA_TIME-SERIE}a), an organized mixed state emerges, forming a checkerboard-like pattern where each ring tends to be in contact with a ring of a different type.

To quantify the level of segregation in the system, we use the parameter $\gamma$, as introduced by Belmonte and collaborators \cite{belmonte2008}. This parameter is defined as the mean fraction of neighboring rings of type 2 (green) surrounding rings of type 1 (red): $\gamma = \left \langle \frac{n_{2}}{n_{1}+n_{2}} \right \rangle_1$, where $\left \langle..\right \rangle_1$ denotes the average over all rings of type 1, with $n_{1}$ and $n_{2}$ representing the number of first neighbors of type 1 and type 2, respectively. In the literature, it is well established that the segregation parameter is directly correlated with the interface $I$ between the types, expressed as $\gamma \sim I$ \cite{beatrici2017,franke2022cell}.

We examined the evolution of the segregation parameter $\gamma$ for various $\Lambda$ values while keeping the $Pe$ value fixed. As shown in Fig. \ref{SNAPSHOTS_and_GAMMA_TIME-SERIE}b), the asymptotic value of $\gamma$ reaches low levels (approximately $\gamma \sim 0.1$ with $10^3$ rings) for the segregated state and increases to $\gamma \sim 0.9$ as $\Lambda$ decreases. The line separating the segregated and checkerboard patterns occurs at $\Lambda \sim 1$, where the initial value of the mixed state ($\gamma \sim 0.7$) is maintained throughout the evolution.

\begin{figure}[!h]
\centering
\includegraphics[width=1.0\columnwidth]{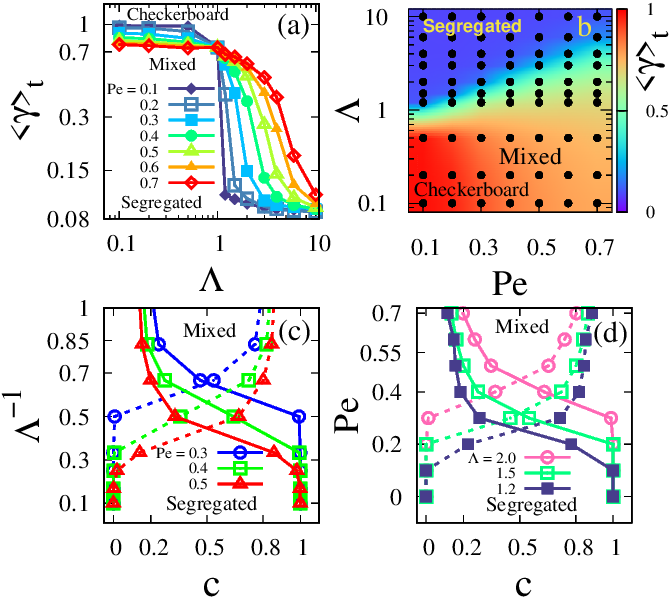}
\caption{(a) Steady-state mean values of segregation parameter $\gamma$ as a function of $\Lambda$ for various values of $Pe$. (b) Diagram ($Pe \times \Lambda$) that separates checkerboard, mixed and  segregated regions. 
Colors represent the average values of segregation parameter for long times.  
(c) Binodal curves from the concentration $c$ of rings of type 1 (red ones) inside the largest mean cluster $c_{in}$ and outside $c_{out}$ (dashed lines) for three values of $Pe$. See text for details. 
(d) Binodal curves from the concentration $c$ of rings of type 1 (red ones) inside the largest mean cluster $c_{in}$ and outside  $c_{out}$ (dashed lines) for three values of $\Lambda$. See text for details.
Fixed parameter: $N = 500$.}
\label{curves_diagrams}
\end{figure}

Furthermore, we measured mean values of steady-state $\gamma$ as a function of $\Lambda$ for several values of $Pe$ (see Fig. \ref{curves_diagrams}a).
 Here we emphasize the role of parameter $Pe$, which broadens the transition as its value increases, and the importance of initial configurations to avoid stay trapped in a meta-stable state. 
While at low $Pe$ values ($Pe \leq 0.2$ in Fig. \ref{curves_diagrams}a) and $\Lambda <1$, the system evolves to the checkerboard pattern, for $\Lambda>1$ it remains trapped close to the initial configuration, so we employed the segregated state as the initial condition to test the effect of $Pe$. 
In brief, $Pe$ functions like temperature, enabling the system to overcome potential barriers to approach the minimum energy state. 
Simultaneously, it is responsible for deviating from the optimal value due to the introduced fluctuations.

In Fig. \ref{curves_diagrams}b we present a $(\Lambda \times Pe)$ diagram, colors indicate the mean value for $\gamma$. 
At low $Pe$ values, the segregation region occurs just above $\Lambda = 1$, but  at high activity we observe a mixed state well above this limit. 
This indicates that the segregation criterion \cite{brodland2002differential, barton2017} $\Lambda_{12} > (\Lambda_{11}+\Lambda_{22})/2$ is only valid at low $Pe$. 
Similar arguments apply to the checkerboard state in the region defined by $\Lambda<1$.

Additionally, we conducted an analysis similar to previous works \cite{weber2016binary} and observed how the steady state of largest cluster of type 1 rings (red) changes with $Pe$ and $\Lambda$.
We define the fraction of these rings that are inside the largest cluster as $c_{in} = {N_{int}/N_{1}}$. $N_1$ and $N_{int}$ are the total number of type 1 rings and the number of rings in the largest asymptotic cluster, correspondingly. 
Using $c_{out} = 1 - c_{in}$, we can construct a binodal curve that delineates regions of segregation and mixing as function of $\Lambda^{-1}$ and $Pe$ (see Figures \ref{curves_diagrams}c-d). 
Below the intersection we have the segregated state ($c_{in}\sim 1$ and $c_{out}\sim 0$, one large cluster) and above, the mixed state (several clusters of similar size). 
Fig. \ref{curves_diagrams}d shows the effect of the activity parameter $Pe$ on the binodal curves. 
Depending on the value of $\Lambda$, there is a  value of $Pe$ beyond which the system transitions from segregated to mixed state.

\begin{figure}[!h]
\centering
\includegraphics[width=1.0\columnwidth]{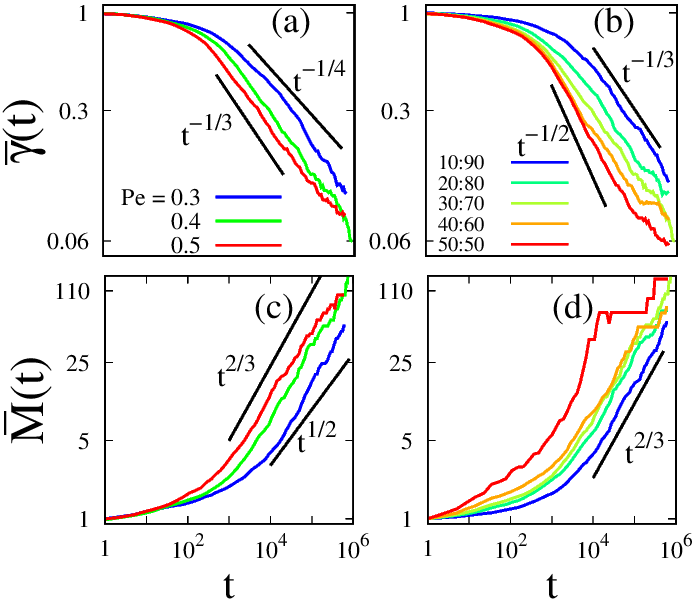}
\caption{
Segregation by differential contraction among rings. (a) Evolution of the rescaled segregation parameter $\bar{\gamma}(t) = \gamma(t)/\gamma_{0}$ and (c) mean cluster size $\bar{M}(t) = M(t)/M_{0}$  for different values of $Pe$. Before reaching saturation, a long-term regime exhibits power-law behavior with an exponent close to $\lambda = -1/3$. Different values of $Pe$ influence the characteristic time at which this power-law regime begins. Ring type ratio: 30:70. 
(b) Temporal evolution of the rescaled segregation parameter $\bar{\gamma}(t)$ and (d) mean cluster size \(\bar{M}(t)\) for different ratios of ring types. Before saturation, the system shows a power-law behavior with an exponent close to $\lambda = -1/3$ and $M \sim \gamma^{-2}$, except for the 50:50 ratio. The activity level is defined as $Pe = 0.4$. The number of rings and the tension in all cases is $N = 10^{4}$ and $\Lambda=10$, respectively.
}
\label{coarsening}
\end{figure}

Our focus now shifts to understanding how the system evolves from a randomly mixed configuration to a segregated one. 
This is particularly important because the long time behavior before saturation of the segregation parameter and the mean size growth of clusters can provide information about the underlying mechanisms occurring during tissue formation. 
The mean cluster's size of rings of type 1, $M(t)$, is obtained using the cluster-counting algorithm used by Beatrici et al.~\cite{beatrici2017}.
Here, we analyze the case where contact between  different rings type is unfavorable ($\Lambda = 10$). 
Initially, the system is in a mixed configuration with a corresponding value of $\gamma\equiv\gamma_0=0.7$, consequent of the (30:70) proportion of ring types. 
In Fig. \ref{coarsening}a, we show the evolution of the segregation parameter normalized by its initial value, $\bar{\gamma}(t) = \gamma(t)/\gamma_0$. 
The activity $Pe$ determines the duration of the transient period, with higher activity implying shorter transients.
Subsequently we observe an algebric decay, $\gamma \sim t^{\lambda}$ with an exponent close to $-1/3$ for at least two decades, independent of the activity value $Pe$. 
Additionally, in this regime, the mean domain size $M$ grows with an exponent close to $2/3$ (see Fig. \ref{coarsening}c). 
Thus, we observe an inverse relationship, $\gamma \sim M^{-2}$, consistent with previous findings \cite{nakajima2011kinetics,krajnc2020solid,durand2021large,franke2022cell}.

In Fig. \ref{coarsening}b, we observe that significantly unequal proportions slow down the domain growth process, as it relies on the coalescence of minority clusters diffusing within the majority type. However, when the proportions are equal (50:50), the dynamics change — the system quickly percolates, and the process of interface reduction becomes dominated by rounding.
In this case, we expect exponent $\sim -1/2$ \cite{Rieu2002}. For other proportions, we observe asymptotic time exponents close to $-1/3$ and $2/3$ for $\gamma$ and $M$, respectively, while maintaining the inverse relationship, $\gamma \sim M^{-2}$ (see Fig. \ref{coarsening}b-d).
 



In conclusion, we present a model where a cell is represented by a ring of active particles. Each ring is explicitly equipped with its own membrane, allowing it to accommodate negative curvatures and replicate all cell sorting mechanisms, whether in isolation or combined. This represents a significant advancement in the class of extended cell models, offering a path to more realistic depiction of cell behavior as observed in experiments.

By employing this comprehensive cellular model and incorporating a term for differential contraction among rings of different types, we conducted a numerical evaluation of Harris' proposed hypothesis on cellular differential surface contraction \cite{harris1976}. 
We found that by keeping all cells with identical attraction forces and solely employing adequate differential interfacial contraction, the system segregates.
We observed a monotonic relationship between the differential contraction parameter $\Lambda$ and the steady-state average value of the segregation parameter $\left\langle \gamma \right\rangle_{t}$, indicating that segregation occurs for $\Lambda > 1$ and a checkerboard pattern emerges for $\Lambda < 1$. 

The rate at which the system transitions from an initially mixed state to its asymptotic state, as well as the ultimate value it reaches, also depends on the activity level $Pe$.
In the case of segregation ($\Lambda>1$), increased activity has a tendency to disrupt the configuration of minimum interfacial energy.
 Binodal curves, showing a transition between the mixed and the segregated states, corroborate the inverse correlation between the contraction interaction $\Lambda$ and the activity $Pe$. 
 These findings suggest that the activity $Pe$ plays a role similar to temperature in thermodynamic equilibrium systems \cite{fily2012athermal,fily2014freezing,bechinger2016active}.
  
  Furthermore, we found that the segregation parameter $\gamma(t)$ and the mean cluster size $M(t)$ exhibit a  power-law regime with an exponent  close to $-1/3$ and $2/3$, respectively. 
  The activity $Pe$ changes the typical timescale at which the asymptotic regime begins but does not modify the growth exponent of the domains. 
Similarly, the proportions do not alter the characteristic exponent, except in the (50:50) case where the system starts close to percolation and the evolution appears to be dominated by rounding. We observed an inverse relationship between the segregation parameter and the typical size, $\gamma \sim M^{-2}$.

Finally, the observed segregation exponent of $\lambda=-1/3$ is at odds with the literature. The cluster-cluster aggregation mechanism seen in our simulations would typically correspond to an exponent of $\lambda=-1/4$, as predicted by surface diffusion models \cite{arenzon2009,cahn1958free} or mean cluster models \cite{beatrici2017mean,kolb84}, particularly in the absence of alignment interactions. An exponent of $\lambda=-1/3$ would instead be expected from the Cahn-Hilliard equation if the mechanism were evaporation-condensation, which we do not observe. The reasons for these discrepancies will be explored in future studies.

We express our gratitude to the Brazilian agencies CAPES, CNPq, and FAPERGS for their financial support. H.C.M.F. and L.G.B. acknowledge the support from the National Council for Scientific and Technological Development – CNPq (procs. 402487/2023-0 and 443517/2023-1). E.F.T. acknowledges ICTP-SAIFR/IFT-UNESP. The simulations were conducted using the \href{https://pnipe.mcti.gov.br/laboratory/19775}{VD Lab} cluster infrastructure at IF-UFRGS.

\section{SUPPLEMENTARY MATERIAL}
\section{A - Boundary Conditions - Repulsive circular walls}
We configure the system as a binary mixture of $N$ active rings confined within a circular arena of radius $R_{0}$. The force exerted by the wall on a specific particle within a ring is given by

\begin{equation}
     \vec{F}_{w} =  -F_{w}\left (\frac{|\vec{r}_{i,j}-\vec{r}_{cm}|}{R_{0}} - 1 \right ) \frac{ ( \vec{r}_{i,j}-\vec{r}_{cm}  )}{|\vec{r}_{i,j}-\vec{r}_{cm}|}
\end{equation}
where $\frac{ ( \vec{r}{i,j}-\vec{r}{cm} )}{|\vec{r}{i,j}-\vec{r}{cm}|}$ is the unit vector connecting the center of particle $i$ within ring $j$ to the center of mass, and $F_{w}$ is the characteristic force exerted by the wall on the particle.
\subsection{B - Forces}

We derive the analytical expressions for the forces acting on each particle $i$ within a ring $j$. These particles are subjected to forces resulting from both shape and interaction energies, which are determined by the area, perimeter, differential line tension, particle-particle overlap, and adhesion terms as defined in the main text. The force on particle $i$ in ring $j$ is obtained by taking the vector derivative of the total energy $E$ with respect to its coordinates.

\begin{eqnarray}
    \vec{r}_{i,j} &=& x_{i,j}\hat{e}_{x} + y_{i,j}\hat{e}_{y}, \ \\ \nonumber \\
     \vec{F}_{i,j} &=& -\frac{\partial E}{\partial \vec{r}_{i,j}} \equiv -\frac{\partial E}{\partial x_{i,j}}\hat{e}_{x}-\frac{\partial E}{\partial y_{i,j}}\hat{e}_{y}.
\end{eqnarray}
\subsubsection{Perimeter force}

The perimeter energy, as defined in the main text for a configuration of $N$ rings labeled by $j = 1, \dots, N$, with $n$ particles labeled $i = 1, \dots, n$, is given by
\begin{eqnarray}
    E_{P} = \frac{\epsilon_{P}}{2} \sum_{j=0}^{N} \sum_{i=0}^{n} \left( \frac{|\vec{l}_{i,j}|}{l_{0}} - 1 \right)^{2}.
\end{eqnarray}
The force on particle $i$ due to deviations in the segment length $|\vec{l}_{i,j}| = |\vec{r}_{i,j} - \vec{r}_{i-1,j}|$ from its preferred value $l_{0}$ is given by
\begin{eqnarray}
    \vec{F}^{i,j}_{P} &=& -\frac{\partial E_{P}}{\partial \vec{r}_{i,j}} = -\frac{\partial E_{P}}{\partial x_{i,j}}\hat{e}_{x} - \frac{\partial E_{P}}{\partial y_{i,j}}\hat{e}_{y}, \\\nonumber\\
      \vec{F}^{i,j}_{P} &=& -\frac{\epsilon_{P}}{l_{0}}\left [ \left (\frac{|\vec{l}_{i,j}|}{l_{0}}-1  \right )\frac{\partial |\vec{l}_{i,j}|}{\partial \vec{r}_{i,j}}  - \left (\frac{|\vec{l}_{i+1,j}|}{l_{0}}-1  \right )\frac{\partial |\vec{l}_{i+1,j}|}{\partial \vec{r}_{i+1,j}} \right ], \nonumber \\\nonumber\\ 
       \vec{F}^{i,j}_{P} &=& \frac{\epsilon_{P}}{l_{0}}\left [ \left (\frac{|\vec{l}_{i+1,j}|}{l_{0}}-1  \right ) \frac{\vec{l}_{i+1,j}}{|\vec{l}_{i+1,j}|} - \left (\frac{|\vec{l}_{i,j}|}{l_{0}}-1  \right )\frac{\vec{l}_{i,j}}{|\vec{l}_{i,j}|} \right ], 
\end{eqnarray}
where we use the following relations:
\begin{eqnarray}
    \frac{\partial |\vec{l}_{i,j}|}{\partial \vec{r}_{i,j}} &=&  \frac{\partial |\vec{l}_{i,j}|}{\partial x_{i,j}}\hat{e}_{x}+  \frac{\partial |\vec{l}_{i,j}|}{\partial y_{i,j}} \hat{e}_{y} = -\frac{\vec{l}_{i,j}}{|\vec{l}_{i,j}|}, \\\nonumber\\
     \frac{\partial |\vec{l}_{i+1,j}|}{\partial \vec{r}_{i+1,j}} &=&  \frac{\partial |\vec{l}_{i+1,j}|}{\partial x_{i+1,j}}\hat{e}_{x}+  \frac{\partial |\vec{l}_{i+1,j}|}{\partial y_{i+1,j}} \hat{e}_{y} = \frac{\vec{l}_{i+1,j}}{|\vec{l}_{i+1,j}|}.
\end{eqnarray}
\subsubsection{Area force}
The force related to area energy on particle $i$ in ring $j$ is given by
\begin{eqnarray}
    \vec{F}^{i,j}_{A}= -\frac{\partial E_{A}}{\partial \vec{r}_{i,j}} = -\frac{\epsilon_{A}}{A_{0}}\left ( \frac{A_{j}}{A_{0}} -1 \right )\frac{\partial A_{j}}{\partial \vec{r}_{i,j}},
\end{eqnarray}
being $E_{A}$ the area energy defined by
\begin{eqnarray}
    E_{A} = \frac{\epsilon_{A}}{2} \sum_{j=0}^{N} \left( \frac{A_{j}}{A_{0}} - 1 \right)^{2}.
\end{eqnarray}
The ring area $A_{j}$ is calculated using the vector product property, given by

\begin{eqnarray}
    A_{j} &=& \frac{1}{2}\sum_{j}^{n} |(\vec{r}_{i,j}-\vec{r}_{cm})\times \vec{l}_{i,j}| \\\nonumber\\
    &=& \frac{1}{2}\sum_{j}^{n} (x_{i,j}-\nonumber x_{cm})(y_{i,j}-y_{i-1,j}) \\\nonumber\\
    &-&(y_{i,j}-y_{cm})(x_{i,j}-x_{i-1,j}),
\end{eqnarray}
where the factor $1/2$ is included to avoid double-counting the area. Therefore, we obtain:

\begin{eqnarray}
     \frac{\partial A_{j}}{\partial x_{i,j}} &=& \frac{\left ( y_{i+1,j} -y_{i-1,j} \right )}{2}, \\ \nonumber\\
      \frac{\partial A_{j}}{\partial y_{i,j}} &=& \frac{\left ( x_{i-1,j} -x_{i+1,j} \right )}{2},
\end{eqnarray}
thus, 
\begin{eqnarray}
    \vec{F}^{i,j}_{A}&=& -\frac{\partial E_{A}}{\partial \vec{r}_{i,j}} = -  \frac{\epsilon_{A}}{2A_{0}}\left ( \frac{A_{j}}{A_{0}} -1 \right ) \\\nonumber\\
     &\times& \left [ \left ( x_{i+1,j} - x_{i-1,j} \right ) \hat{e}_{x} - \left ( y_{i+1,j} - y_{i-1,j} \right )\hat{e}_{y}   \right ]. \nonumber
\end{eqnarray}
\subsubsection{Interaction forces}

The core repulsion for non-neighboring particles within the same ring and the adhesion between particles of different rings are accounted for by the interaction energy $E_{int}$, which is described by a truncated harmonic potential,

\begin{eqnarray}
\displaystyle E_{int} &=& \left\{
\begin{array}{lcl}
\displaystyle \frac{\epsilon_{c}}{2}\left( \frac{r_{ik}}{\sigma} - 1 \right)^{2}, & & r_{ik} \leq  \sigma \\[10pt]
\displaystyle \frac{\epsilon_{adh}}{2}\left( \frac{r_{ik}}{\sigma} - 1 \right)^{2}, & &  l_{adh} \geq r_{ik} > \sigma \\[10pt]
\displaystyle 0, & & r_{ik} > l_{adh}.
\end{array}
\right.
\end{eqnarray}
Therefore, the corresponding force
\begin{eqnarray}
    \vec{F}^{i,j}_{int} &=& -\frac{\partial E_{int}}{\partial \vec{r}_{i,j}}, \\\nonumber\\
    \vec{F}^{i,j}_{int} &=& \hat{r}_{ik}\left\{
\begin{array}{lcl}
\displaystyle -\frac{\epsilon_{c}}{\sigma}\left( \frac{r_{ik}}{\sigma} - 1 \right), & & r_{ik} \leq  \sigma \\[10pt]
\displaystyle -\frac{\epsilon_{adh}}{\sigma}\left( \frac{r_{ik}}{\sigma} - 1 \right), & &  l_{adh} \geq r_{ik} > \sigma \\[10pt]
\displaystyle 0, & & r_{ik} > l_{adh}
\end{array}
\right.
\end{eqnarray}
where $\hat{r}_{ik} = \frac{\vec{r}_{ik}}{r_{ik}}$ is a unit vector connecting particle $i$ and $k$.
\subsubsection{Differential line tension (contraction)}
The contraction energy represents a line tension that acts between two particles from different rings when they are within a distance cutoff $l_{\Lambda}$. This energy term is given by

\begin{eqnarray}
    E_{\Lambda} =  \sum_{j=0}^{N} \sum_{i=0}^{n} \Lambda_{\alpha \beta} |\vec{l}_{i,j}|.
\end{eqnarray}
The force acting on particle $i$ inside ring $j$ is
\begin{eqnarray}
     \vec{F}^{i,j}_{\Lambda} &=& -\frac{\partial E_{\Lambda}}{\partial \vec{r}_{i,j}}, \\ \nonumber\\
     \vec{F}^{i,j}_{\Lambda} &=& \Lambda_{\alpha \beta}\left [  \frac{\vec{l}_{i+1,j}}{|\vec{l}_{i+1,j}|} - \frac{\vec{l}_{i,j}}{|\vec{l}_{i,j}|} \right ]. 
\end{eqnarray}
Therefore, the force $\vec{F}^{i,j}_{\Lambda}$ on particle $i$ within ring $j$ depends on the type $\alpha$ of its ring and the type $\beta$ of the ring of the particle it interacts with. If the particle is within the cutoff distance and interacts with multiple particles from other rings, the tension value will be the mean of all the individual tensions. Specifically, this mean is given by
\begin{eqnarray}
    \Lambda_{\alpha \beta} = \frac{1}{n_c}
\sum_{k}^{}\Lambda_{\alpha k},
\end{eqnarray}
where the sum $k$ runs over all contacts (both $\alpha-\alpha$ and $\alpha-\beta$ interactions) and $n_{c}$ represents the total number of contacts.

\subsection{C - Control parameters}

The value of the total equilibrium area $A_{r}$ is the sum of the equilibrium area imposed by the area elastic energy term   plus half the area of each particle composing the ring, 
 \begin{eqnarray}
 A_{r} = A_{0} + \frac{n\pi \sigma^{2}}{8}. 
 \end{eqnarray}
We define a packing fraction $\phi$ for the system, relative to a circular region 
of radius $R_0$,  through the relation 
\begin{eqnarray}
   \phi = \frac{NA_{r}}{\pi R_{0}^{2}}. 
\end{eqnarray}
We fix $\phi =  0.895$  allowing for neighbor exchange and neighbor interaction simultaneously and we use a fixed dimensionless perimeter-area equilibrium ratio $p_{0} = P_{0}/\sqrt{A_{0}} = 4$, where $P_{0} = nl_{0}$ is the equilibrium perimeter. 
The shape parameter $p_{0}$ defines the degree of stiffness of the ring. 
The choice of this value is based on previous works with Vertex \cite{bi2015density,barton2017,krajnc2020solid}, Voronoi \cite{bi2016motility}, and ring models \cite{boromand2018jamming}, where an excess of cell perimeter is found for $p_{0} > 3.81$ as well as the emergence of a liquid-like behavior.  To prevent overlap among rings, we reached a compromise by setting comparable values for $\epsilon_c$, $\epsilon_P$ and $\epsilon_A/n$, while assigning a much lower value to $\epsilon_{adh}$. 
So, we simulate the ring system keeping the following parameters fixed:  $\epsilon_{c}/\epsilon_{P} = 1$, $\epsilon_{A}/\epsilon_{P} = 35$, $(F_{w}\sigma)/\epsilon_{P} = 1$, $\epsilon_{adh}/\epsilon_{P} = 5.10^{-4}$, $n = 10$, $l_{0}= 1$, $\sigma = l_{0}$, $l_{\Lambda} = l_{adh} = 1.5\,l_{0}$ and $\mu = 1$. 
This choice of parameters ensures that $\epsilon_{P}$ and $\epsilon_{A}$ are sufficiently large to maintain the bond length close to $l_{0}$ and the equilibrium area close to $A_{0}$.  The adhesion forces between different rings are kept equal for all rings. 
Furthermore, to emphasize the effects of differential contractions, we adopted an adhesion value that is considerably lower compared to the other energy terms involved. We integrate the equations of motion, using the Euler-Maruyama algorithm with a time step $\Delta t = 0.01$.
 
\subsection{D - Neighbors criterion for measurements}
In this work, we use the criterion for defining whether two rings are neighbors based on the distance between their centers of mass, $d_{jk} = |\vec{R}_{j} - \vec{R}_{k}| \leq \frac{P_{0}}{2} = 5\sigma$. This relation ensures that even two completely flattened rings in contact will be considered neighbors.

\bibliographystyle{apsrev4-2}
\bibliography{multi_ring}

\end{document}